\documentclass{elsart}
\usepackage{graphicx,latexsym,amssymb}
\newcommand{\dps}{\displaystyle}

\journal{Chemical Physics}

\begin{document}
\begin{frontmatter}
\title{An Initial Value Representation with Complex Trajectories}

\author{Marcus A.M. de Aguiar\corauthref{cor}}
\corauth[cor]{Corresponding author.}
\ead{aguiar@ifi.unicamp.br}
\author{Silvio A. Vitiello}
\author{ and}
\author{Adriano Grigolo }

\address{$^1$Instituto de F\'{\i}sica Gleb Wataghin, Universidade Estadual de
Campinas, UNICAMP, 13083-970 Campinas, S\~ao Paulo, Brasil}

%%%%%%%%%%%%%%%%%%%%%%%%%%%%%%%%%%%%%%%%%%%%%%%%%%%%%%%%%%%%%%%%%%%%%
\begin{abstract}

We present an Initial Value Representation for the semiclassical
coherent state propagator based on complex trajectories. We map the
complex phase space into a real phase space with twice as many
dimensions and use a simple procedure to automatically eliminate
non-contributing trajectories. The resulting semiclassical formulas do
not show divergences due to caustics and provide accurate results.

\end{abstract}

\begin{keyword}
semiclassical methods; coherent states; initial value representations;
complex trajectories

%\PACS{03.65.Sq, 31.15.Gy, 05.45.Mt}
% 03.65.Sq Semiclassical theories and applications
% 31.15.Gy Semiclassical methods
% 05.45.Mt 	Quantum chaos; semiclassical methods

\end{keyword}
\end{frontmatter}

%%%%%%%%%%%%%%%%%%%%%%%%%%%%%%%%%%%%%%%%%%%%%%%%%%%%%%%%%%%%%%%%%%%%%
%%%%%%%%%%%%%%%%%%%%%%%%%%%%%%%%%%%%%%%%%%%%%%%%%%%%%%%%%%%%%%%%%%%%%
\section{Introduction}

The investigation of wavepackets dynamics by semiclassical methods has
practical importance for calculations of several processes involving
atoms and molecules. It is also a fundamental topic in the study of the
classical-quantum connection, especially for chaotic systems and for
open systems, coupled to environments.

The history of semiclassical methods goes back to the origins of
quantum mechanics itself. One fundamental result is the so called Van
Vleck approximation to the coordinate propagator, derived in 1928
\cite{Van28}, that can be written as
\begin{equation}
   \label{vv}
\langle x_f|e^{-i\hat{H}\tau/\hbar}|x_i\rangle \approx
   \frac{1}{\sqrt{2\pi m_{qp}}} \;  e^{i S(x_i,x_f,T) /\hbar - i\pi/4- i\pi k/2} \; .
\end{equation}
In this expression $S(x_i,x_f,T)$ is the classical action of a
trajectory connecting coordinates $x_i$ to $x_f$ in the time $T$,
$m_{qp}$ is an element of the tangent matrix, that controls the motion
in the vicinity of this trajectory, and $k$ is the number of focal
points (where $m_{qp}$ goes to zero) along the trajectory. If more than
one trajectory satisfying these boundary conditions exists, one has to
sum their contributions. From this basic propagator one can compute the
time evolution of arbitrary wavefunctions.

A more direct approach to calculate the time evolution of wavepackets
in given by the propagator in the coherent state representation. The
coherent states of the harmonic oscillator are minimum uncertainty
wavepackets and define a representation involving both the coordinates
and the momenta that can be readily visualized in the phase space. The
coherent state propagator $\langle z_f | e^{-\frac{i }{\hbar}\hat{H}T}
| z_0 \rangle$ represents the amplitude probability that the initial
wavepacket $| z_0 \rangle$ centered on $q_0,p_0$ is found at the state
$| z_f \rangle$, centered on $q_f,p_f$, after a time $T$. However, the
direct evaluation of the semiclassical limit of this propagator results
in an expression bearing the same difficulties of the Van Vleck formula
\cite{Klau85,Bar01,eva07,garg07}, namely: (a) the classical
trajectories needed are defined by mixed initial-final boundary
conditions, rendering the calculation hard, specially in
multidimensional or chaotic systems and; (b) the formula diverges at
phase space focal points. Moreover, the trajectories are complex and
some of them, even satisfying the appropriate boundary conditions, lead
to unphysical contributions and must be discarded
\cite{Hel87,Hel88,adachi,Rub95,Shu95,Shu96,Rib04,Agu05}.

Several methods have been developed to overcome these difficulties,
most of them based on the idea of initial value representations (IVR)
\cite{Mill70,Mill74,Hell75,Herm84,Kay94a,Kay94b,Kay97,Mill01,Zha03,Zha04}.
 Among
these, the Herman-Kluk propagator \cite{Herm84} and the method of
linearized cellular dynamics developed by Heller and Tomsovic
\cite{Hell91,Tom91} stands out as very accurate. More recent
derivations and corrections to the basic Herman-Kluk formula have also
provided new insight into this class of approximation
\cite{dimitri,dim08,Pol03,Kay06}.

In spite of the many difficulties involved in the calculation of the
coherent state propagator with complex trajectories, this approximation
turns out to be very accurate \cite{adachi,Rub95,Shu95,Hell02,Rib04}.
Recent work on Bohmian mechanics have also employed complex
trajectories, providing a new formulation leading to accurate results
\cite{Tan06,Tan07}. In this paper we propose the construction of an
initial value representation for this approximation that removes most
of its problems: the mixed conditions defining the trajectories are
replaced by initial conditions; the complex trajectories are mapped
into real trajectories of an associated Hamiltonian; the divergences
due to focal points are eliminated and; a simple and automatic
filtering is used to eliminate the non-contributing trajectories.

This paper is organized as follow. The next section reviews the
semiclassical coherent state propagator and its semiclassical
approximation in terms of complex trajectories. In section
\ref{sec_CIVR} we develop the initial value representation for complex
trajectories (CIVR). It presents what we call the sudden CIVR, where
only trajectories satisfying the original mixed conditions are
considered, and the smooth CIVR where the neighborhood of the relevant
trajectories are considered as well. We also discuss how the complex
trajectories calculations are performed in terms of real trajectories.
This section ends with a discussion of the criteria used for filtering
out the non-contributing trajectories. In section \ref{sec_Ex} the
smooth CIVR is applied to an anharmonic quartic oscillator and some
final comments are made. In two appendices useful relations of the
tangent matrix are derived for both real and complex trajectories.

%%%%%%%%%%%%%%%%%%%%%%%%%%%%%%%%%%%%%%%%%%%%%%%%%%%%%%%%%%%%%%%%%%%%%
%%%%%%%%%%%%%%%%%%%%%%%%%%%%%%%%%%%%%%%%%%%%%%%%%%%%%%%%%%%%%%%%%%%%%
\section{The semiclassical coherent state propagator}

The coherent state $|z\rangle$ of a harmonic oscillator of mass $m$ and
frequency $\omega$ is defined by
\begin{equation}
  \label{cs}
  |z\rangle = e^{-\frac{1}{2}|z|^2}e^{z\hat{a}^\dagger}|0\rangle
\end{equation}
with $|0\rangle$ the harmonic oscillator ground state and
\begin{equation}
  \label{aad}
  \hat{a}^\dagger = \frac{1}{\sqrt{2}}\left( \frac{\hat{q}}{b}-i
              \,\frac{\hat{p}}{c} \right), \qquad
  z =  \frac{1}{\sqrt{2}}\left( \frac{q}{b}+i
              \,\frac{p}{c} \right).
\end{equation}
In these equations $\hat q$, $\hat p$, and $\hat{a}^\dagger$ are
operators; $q$ and $p$ are real numbers and $z$ is complex.  The
parameters $b = {(\hbar/ m \omega )}^{\frac{1}{2}}$ and $c = {(\hbar m
\omega )}^{\frac{1}{2}}$ define the length and momentum scales,
respectively, and their product is $\hbar$.

For a time-independent Hamiltonian operator $\hat{H}$, the propagator
in the coherent states representation is the matrix element of the
evolution operator between states $|z_0 \rangle$ and $| z_f \rangle$
\cite{Kla87}:
\begin{equation}
  \label{csp}
  K(z_f^*,z_0,T) = \langle z_f | e^{-\frac{i }{\hbar}\hat{H}T}
                  | z_0 \rangle.
\end{equation}

The semiclassical evaluation of $K(z_f^*,z_0,T)$ was presented in
detail in \cite{Bar01,eva07}. The result is given by
\begin{eqnarray}
  \label{scsp}
  K_{sc}(z_f^*,z_0,T) = \sum_\nu
         \sqrt{\frac{i}{\hbar}\frac{\partial^2 S}
         {\partial z_0 \partial z_f^*}} \;
         \exp \left\{ \frac{i}{\hbar}(S+I) - \frac{1}{2}
          \bigl( |z_f|^2 + |z_0|^2\bigr) \right\} \, ,
\end{eqnarray}
where
\begin{eqnarray}
  \label{cact}
  S = S(z_f^*,z_0,t)
         &=\int\limits_0^t d t' \left[\frac{i \hbar}{2}
          (\dot{u}v-\dot{v}u) - H(u,v,t') \right]
          - \frac{i \hbar}{2} ( u(T)z_f^* + z_0 v(0) )
\end{eqnarray}
is the action and the classical Hamiltonian function is calculated from
the Hamiltonian operator as $H(u,v) = \langle v|\hat{H}| u \rangle$.
The term
\begin{equation}
\label{ifac}
I = \frac{1}{2} \int_0^T
\frac{\partial^2 H}{\partial u \partial v}
{\rm d}t
\end{equation}
is a correction to the action. The sum over $\nu$ represents the sum
over all contributing (complex) classical trajectories satisfying
Hamilton's equations with boundary conditions
\begin{equation}
\label{bounda} \frac{1}{\sqrt{2}}\left( \frac{q(0)}{b}+i
              \,\frac{p(0)}{c} \right) = z_0~,
              \qquad  \frac{1}{\sqrt{2}}\left( \frac{q(T)}{b}-i
              \,\frac{p(T)}{c} \right) = z_f^\star\;.
\end{equation}

In all these expressions the variables $u$ and $v$ are defined by
\begin{equation}
  \label{uveq}
u = \frac{1}{\sqrt{2}}\left(\frac{q}{b}
       +i \,\frac{p}{c} \right), \qquad \qquad
v = \frac{1}{\sqrt{2}}\left(\frac{q}{b}
       -i \,\frac{p}{c} \right).
\end{equation}
They are manifestly independent ($u \neq v^*$ since $q$ and $p$ are
complex), and replace $z$ and $z^*$ to avoid confusion. In these
variables the boundary conditions become
\begin{equation}
\label{bound} u(0) = z_0~, \qquad v(T) = z_f^\star\;.
\end{equation}
%

%%%%%%%%%%%%%%%%%%%%%%%%%%%%%%%%%%%%%%%%%%%%%%%%%%%%%%%%%%%%%%%%%%%%%
%%%%%%%%%%%%%%%%%%%%%%%%%%%%%%%%%%%%%%%%%%%%%%%%%%%%%%%%%%%%%%%%%%%%%
\section{A complex initial value representation}
\label{sec_CIVR}

%%%%%%%%%%%%%%%%%%%%%%%%%%%%%%%%%%%%%%%%%%%%%%%%%%%%%%%%%%%%%%%%%%%%%
\subsection{Basic idea}

The first of the boundary conditions (\ref{bound}) specifying the
complex trajectory can be written explicitly as
\begin{equation}
\frac{q(0)}{b} + i b\frac{p(0)}{\hbar} = \frac{q_0}{b} + i b\frac{p_0}{\hbar} ,
\label{civr1}
\end{equation}
where $q_0$ and $p_0$ define the initial coherent state $|z_0\rangle$.
This condition is not sufficient to determine the trajectory, since
$q(0)$ and $p(0)$ are complex. The missing condition is given by the
second equation in (\ref{bound}) and refers to the final propagation
time $T$.

In order to avoid dealing with mixed initial-final conditions, let us
first suppose we have had a second
equation of the form
\begin{equation}
\frac{q(0)}{b} - i b \frac{p(0)}{\hbar} = \frac{q_1}{b} - i b\frac{p_1}{\hbar} .
\label{civr3}
\end{equation}
By solving for $q(0)$ and $p(0)$ one finds
\begin{equation}
\begin{array}{ll}
q(0) &= \frac{1}{2}\left[(q_0+q_1) + i \frac{b^2}{\hbar}(p_0 - p_1)\right]   \\
p(0) &= \frac{1}{2}(\left[p_0+p_1) + i \frac{\hbar}{b^2}(q_1 - q_0)\right].
\end{array}
\label{civr4}
\end{equation}
For $q_0$ and $p_0$ fixed, each $q_1$ and $p_1$ defines a trajectory
with end points $q(T)$ and $p(T)$.

Let $\tilde{q}_1$ and $\tilde{p}_1$ be the values of $q_1$ and $p_1$
such that the second of equations (\ref{bound}) is satisfied, i.e., for
which the initial conditions (\ref{civr4}) leads to
\begin{equation}
\frac{q(T)}{b} - i b\frac{p(T)}{\hbar} = \frac{q_f}{b} - i b\frac{p_f}{\hbar},
\label{civr5}
\end{equation}
where $q_f$ and $p_f$ define the final coherent state $|z_f\rangle$.
Then, we can rewrite the semiclassical propagator of Eq.~(\ref{scsp})
as
\begin{eqnarray}
  \label{scspi}
  K_{sc}(z_f^*,z_0,T) & = & \int dq_1 dp_1 \delta_a(q_1 -\tilde{q}_1)
        \delta_a(p_1 -\tilde{p}_1) \nonumber\\
         & \times & \sqrt{\frac{i}{\hbar}\frac{\partial^2 S}
         {\partial z_0 \partial z_f^*}} \;
         e^{\frac{i}{\hbar}(S+I) - \frac{1}{2}
          \bigl( |z_f|^2 + |z_0|^2\bigr)},
\end{eqnarray}
where the trajectories are now calculated according to the initial
conditions (\ref{civr4}) and their contributions filtered out by the
delta functions. Since these are sharped peaked functions, equation
(\ref{scspi}) is identical to (\ref{scsp}), because only the
trajectories satisfying the proper boundary conditions (\ref{bound})
are taken into account. If the delta functions are replaced by
Gaussian functions of width $a$, Fillinov type expansions become possible and a
smoothed and better behaved expression arises.

The equivalent expression of the semiclassical propagation for an
arbitrary initial state described by the  wave-function $\psi(z_0^*,0)
= \langle z_0|\psi\rangle$ is
\begin{eqnarray}
  \label{scspig}
  \psi(z_f^*,T) = \frac{1}{2\pi\hbar} \int  K_{sc}(z_f^*,z_0,T)
            \psi(z_0^*,0)
             \delta_a(q_1 -\tilde{q}_1)
            \delta_a(p_1 -\tilde{p}_1) \, dq_0 dp_0 dq_1 dp_1.
\end{eqnarray}

We note that the second derivative of the action with respect to its
arguments, as appearing in the pre-factor of the semiclassical
propagator, can be written in terms of the tangent matrix, that
controls the classical motion in the vicinity of a given trajectory. In
appendix \ref{appA} we derive several useful relations between the
tangent matrix in $u$, $v$ and $q$, $p$ variables for complex and real trajectories.
In particular, we show that
\begin{equation}
\frac{i}{\hbar} \frac{\partial^2 S}{\partial z_0 \partial z_f^*} = \frac{1}{M_{vv}}.
\end{equation}

Before we end this subsection we define the scaled coordinates and
momenta $\bar{q} = q/b$ and $\bar{p} = p b/\hbar$. Defining the scaled
Hamiltonian
\begin{equation}
\bar{H}(\bar{q},\bar{p}) = \frac{1}{\hbar} H(b \bar{q}, \hbar \bar{p}/b)
\label{barvar}
\end{equation}
it is easy to check that the semiclassical expressions in terms of
$\bar{q}$, $\bar{p}$ and $\bar{H}$ become identical to the original
expressions with $b$  and $\hbar$ replaced by 1. Therefore, from now
one we shall use these scaled variables, which amounts to set
$b=\hbar=1$, but will omit the bar to make the notation simpler. The
original variables will be recovered later in the examples.

%%%%%%%%%%%%%%%%%%%%%%%%%%%%%%%%%%%%%%%%%%%%%%%%%%%%%%%%%%%%%%%%%%%%%
\subsection{The calculation of complex trajectories}
\label{sec_CCT}

For analytic Hamiltonian functions $H(q,p)$ it is possible to rewrite
the equations of motion for the complex variables $q$ and $p$ in terms
of real trajectories of an auxiliary Hamiltonian system with twice as
many degrees of freedom, or as we call it, the double phase space.
The definitions \cite{xav96,kau00}
\begin{eqnarray}
q = Q_1 + i P_2, \qquad \qquad p=P_1+iQ_2
\label{varnew}
\end{eqnarray}
and
\begin{eqnarray}
  \label{hnew}
H(q,p) = H_1(Q_1,Q_2,P_1, P_2) + i H_2(Q_1,Q_2,P_1, P_2),
\end{eqnarray}
where $H_1$ and $H_2$ are real functions, allows to show easily that
Hamilton's equations for $q$ and $p$ are equivalent to
\begin{eqnarray}
  \label{eqnew}
\dot{Q}_i = \frac{\partial H_1}{\partial P_i} , \qquad \qquad
\dot{P}_i = -\frac{\partial H_1}{\partial Q_i}, \qquad \qquad i=1,2.
\end{eqnarray}
Note that $H_2$ is also a constant of the motion. The separation of
variables in (\ref{varnew}) may look unusual because it mixes q's and
p's, but this is the proper combination to get the correct signs in
Hamilton's equations. These separation of variables also look natural
when the form of equation (\ref{civr4}) is considered.

For the case  $|\psi(0)\rangle = |z_0\rangle$, the real trajectory
starting from the center of the wavepacket plays an important role, and
we shall use it as a reference. Therefore the integration over $q_1$
and $p_1$ in the CIVR will be centered on $q_0$ and $p_0$ and only a
limited region around this point is expected to significantly
contribute to the propagation. In this way we write
\begin{eqnarray}
  \label{ininew}
q_1 = q_0 + \Delta q, \qquad \qquad p_1 = p_0 + \Delta p
\end{eqnarray}
and the initial conditions (\ref{civr4}) reduce to
\begin{equation}
\begin{array}{ll}
Q_1(0) &= q_0 + \Delta q/2  \qquad \qquad Q_2(0) = \Delta q/2  \\
P_1(0) &= p_0 + \Delta p/2  \qquad \qquad P_2(0) = -\Delta p/2.
\end{array}
\label{civr4n}
\end{equation}
In accordance with Eq.~(\ref{varnew}), $q(0)=q_0+w$, $p(0)=p_0+iw$ with
$w=(\Delta q - i \Delta p)/2$, which is exactly the variable used in a
search procedure developed in ref.\cite{Rub95}.

All the tangent matrix elements appearing in equations (\ref{smivrg})
and (\ref{smivrag}) can be readily computed from the tangent matrix of
the real trajectory in the double phase space. This procedure
eliminates the need to work with complex trajectories and also the so
called root search problem, involved in finding trajectories with mixed
initial-final conditions.

%%%%%%%%%%%%%%%%%%%%%%%%%%%%%%%%%%%%%%%%%%%%%%%%%%%%%%%%%%%%%%%%%%%%%
\subsection{The connection between initial and final displacements}

The connection between the initial and final displacements can be
established as follow. Initially by
comparing equations (\ref{varnew}) with (\ref{civr4}) we see that
\begin{equation}
\begin{array}{ll}
Q_1(0) &= \frac{1}{2}(q_0+q_1)  \\
Q_2(0) &= \frac{1}{2}(q_1 - q_0) \\
P_1(0) &= \frac{1}{2}(p_0+p_1) \\
P_2(0) &= \frac{1}{2}(p_0 - p_1)   \\
\end{array}
\label{civr10}
\end{equation}
which also leads to $q_1 = Q_1(0)+Q_2(0)$ and $p_1=P_1(0)-P_2(0)$. It
turns out to be convenient to extend this definition to
\begin{equation}
\begin{array}{ll}
q_1(t) = Q_1(t)+Q_2(t)  \\
p_1(t) =P_1(t)-P_2(t).  \\
\end{array}
\label{civr11}
\end{equation}

Because of the filtering functions in (\ref{scspi}) and (\ref{scspig})
(smoothed or sharp) the relevant contributions to the integrals over
$q_1$ and $p_1$ come from the vicinities of $\tilde{q}_1$ and
$\tilde{p}_1$, that should also be close to $q_0$ and $p_0$. For this
particular trajectory $v(T)=z_f^*$:
\begin{equation}
[Q_1(T)+iP_2(T)] -i [P_1(T) + i Q_2(T)] = q_f - i q_f
\label{civr12}
\end{equation}
or, according to (\ref{civr11}), $q_1(T)=q_f$ and $p_1(T)=p_f$.

For neighboring trajectories we may expand the final values of $q_1(T)$
and $p_1(T)$ around $q_f$ and $p_f$ as:
\begin{displaymath}
\begin{array}{ll}
q_1(T) &\approx q_f + \frac{\partial q_1(T)}{\partial q_1}(q_1-\tilde{q}_1) +
                    \frac{\partial q_1(T)}{\partial p_1}(p_1-\tilde{p}_1) \\
p_1(T) &\approx p_f + \frac{\partial p_1(T)}{\partial q_1}(q_1-\tilde{q}_1) +
                    \frac{\partial p_1(T)}{\partial p_1}(p_1-\tilde{p}_1)
\end{array}
\end{displaymath}
or
\begin{equation}
\left( \begin{array}{l}
    q_1(T) - q_f \\
    p_1(T) - p_f \end{array} \right) =
\left( \begin{array}{cc}
    \displaystyle{\frac{\partial q_1(T)}{\partial q_1}} \quad &
    \displaystyle{\frac{\partial q_1(T)}{\partial p_1}}  \\
    \displaystyle{\frac{\partial p_1(T)}{\partial q_1}} \quad &
    \displaystyle{\frac{\partial p_1(T)}{\partial p_1}}
    \end{array} \right)
\left( \begin{array}{l}
    q_1-\tilde{q}_1 \\
    p_1-\tilde{p}_1 \end{array} \right) \equiv \Lambda
\left( \begin{array}{l}
    q_1-\tilde{q}_1 \\
    p_1-\tilde{p}_1 \end{array} \right).
\label{civr6}
\end{equation}
It follows that
\begin{equation}
\delta(q_1 -\tilde{q}_1) \delta(p_1 -\tilde{p}_1) = |\det{\Lambda} | ~
\delta(q_1(T) -q_f) \delta(p_1(T) -p_f).
\label{civr7}
\end{equation}
In appendix \ref{appB}, equation (\ref{b6}), we show that
$\det{\Lambda} = |M_{vv}|^2$.

%%%%%%%%%%%%%%%%%%%%%%%%%%%%%%%%%%%%%%%%%%%%%%%%%%%%%%%%%%%%%%%%%%%%%
\subsection{Sudden complex initial value representation}

In the case of sharp delta functions we can use equation (\ref{civr7})
to write down the first of our formulas, that we term {\it sudden}
CIVR. Since
\begin{equation}
\frac{1}{\sqrt{2}}\left(q_1(T) -i \,p_1(T) \right) = v(T)
\label{eqvt}
\end{equation}
and defining
\begin{equation}
\delta^2(v(T) - z_f^*) = 2 \pi \delta(q_1(T) -q_f) \delta(p_1(T) -p_f)
\end{equation}
we obtain
\begin{eqnarray}
  \label{sivr}
  \psi(z_f^*,T) = \int |M_{vv}|^{3/2}
         e^{i(S+I) - \frac{1}{2} \bigl( |z_f|^2 + |z_0|^2\bigr) - i\frac{\xi}{2}}
            \psi(z_0^*,0) \, \delta^2(v(T) - z_f^*) \frac{d^2 z_0}{\pi} \frac{d^2 v_1}{\pi}
\end{eqnarray}
where $\xi$ is the phase of $M_{vv}$. Each pair of phase space points
$q_0$, $p_0$ and $q_1$, $p_1$ define a complex trajectory with initial
conditions
\begin{equation}
\begin{array}{ll}
q(0) &= \frac{1}{2}(q_0+q_1) + i \frac{1}{2}(p_0 - p_1)   \\
p(0) &= \frac{1}{2}(p_0+p_1) + i \frac{1}{2}(q_0 - q_1).
\end{array}
\label{civr4a}
\end{equation}
The contribution of these trajectories to the final result are filtered
by the delta function. The integration measures are defined as usual as
$d^2z_0/\pi = dq_0 dp_0/2\pi$ and $d^2v_1/\pi = dq_1 dp_1/2\pi$.

Notice that the arguments of $S$ and $I$ in (\ref{sivr}), which were
originally $(z_f^*,z_0,T)$, can be replaced by $(v(T),z_0,T)$, so that
both $S$ and $I$ are computed for the trajectories defined by
(\ref{civr4a}). Also important is the fact that $M_{vv}$ in the
pre-factor has moved from the denominator to the nominator  so that
divergences at caustics are replaced by non-contributing trajectories.
This is a well known property of IVR's constructed in this way.

%%%%%%%%%%%%%%%%%%%%%%%%%%%%%%%%%%%%%%%%%%%%%%%%%%%%%%%%%%%%%%%%%%%%%
\subsection{Smooth complex initial value representation}

If the delta functions in the CIVR are replaced by Gaussian functions a
more well behaved approximation is obtained. Following Filinov
\cite{Fil86} and Makri \cite{Mak87} we replace the filtering integrals
of trajectories according to
\begin{eqnarray}
\int \delta(v_1 - \tilde{v}) \frac{d^2 v_1}{\pi} & \rightarrow &
\int e^{-\frac{1}{2a^2}[(q_1 - \tilde{q}_1)^2 +(p_1 - \tilde{p}_1)^2]}
\frac{d^2 v_i}{\pi a^2} \nonumber \\
& \approx &
\int e^{-\frac{1}{2a^2 |M_{vv}|^2}[(q_1(T) - q_f)^2 +(p_1(T) - p_f)^2]}
\frac{d^2 v_i}{\pi a^2} \nonumber \\
& = & \int |M_{vv}|^2 e^{-\frac{|v(T)-z_f^*|^2}{\alpha^2}} \frac{d^2 v_1}{\pi \alpha^2},
\label{gau}
\end{eqnarray}
where we have used equation (\ref{civr6}) in the second line and
defined the re-scaled width
\begin{equation}
\alpha = a |M_{vv}|.
\label{alpha}
\end{equation}

The use of smooth filters seems appropriate to coherent state
propagation. It implies that not only the trajectories satisfying the
exact boundary conditions (\ref{bound}) are considered, but also their
neighborhood as defined by the parameter $a$. In this case the action
$S(z_f^*,z_0,T)$ in equation (\ref{scspi}) cannot be simply replaced by
$S(v(T),z_0,T)$, but has to be expanded around each initial value
trajectory up to second order. The result is
\begin{eqnarray}
S(z_f^*,z_0,T) &\approx & S(v(T),z_0,T) + \frac{\partial S}{\partial v(T)}(z_f^*-v(T)) +
\frac{1}{2} \frac{\partial^2 S}{\partial v(T)^2}(z_f^*-v(T))^2 \nonumber \\
& \approx & S(v(T),z_0,T) - i u(T)(z_f^*-v(T)) -i \frac{M_{uv}}{2 M_{vv}}(z_f^*-v(T))^2,
\label{actexp}
\end{eqnarray}
where once again we have resorted to expressions derived in appendix
\ref{appB}.

The {\it smooth} CIVR can then be obtained by using equations
(\ref{gau}) and (\ref{actexp}) in (\ref{scspig}):
\begin{eqnarray}
  \label{smivr}
  \psi(z_f^*,T) = \int |M_{vv}|^{3/2}
         \exp{ \left\{ \phi-\frac{\left|v(T)-z_f^*\right|^2}{\alpha^2} \right\} }
         \psi(z_0^*,0) \, \frac{d^2 z_0}{\pi} \frac{d^2 v_1}{\pi\alpha^2},
\end{eqnarray}
where
\begin{eqnarray}
  \label{smivra}
   \phi  =  i(S + I) + u(T)\left(z_f^*-v(T)\right) + \frac{M_{uv}}
            {2 M_{vv}}\left(z_f^*-v(T)\right)^2 - \frac{|z_f|^2}{2} - \frac{|z_0|^2}{2}
    - i\frac{\xi}{2} .
\end{eqnarray}

If the initial state to be propagate is itself a coherent state,
equation (\ref{scspi}), the smooth CIVR simplifies to
\begin{eqnarray}
  \label{smivrg}
  K(z_f^*,z_0,T) = \int |M_{vv}|^{3/2}
         \exp{ \left\{ \phi-\frac{\left|v(T)-z_f^*\right|^2}{\alpha^2} \right\} }
          \, \frac{d^2 v_1}{\pi \alpha^2},
\end{eqnarray}
with
\begin{eqnarray}
  \label{smivrag}
   \phi & = & i(S + I) + u(T) z_f^* + \frac{M_{uv}}
            {2 M_{vv}}\left(z_f^*-v(T)\right)^2 - \frac{|z_f|^2}{2}
            - \frac{|z_0|^2}{2} - i\frac{\xi}{2} .
\end{eqnarray}
In this paper we shall discuss an example of this simple case only.

%%%%%%%%%%%%%%%%%%%%%%%%%%%%%%%%%%%%%%%%%%%%%%%%%%%%%%%%%%%
\subsection{Filtering out non-contributing trajectories}

It is well known that not all trajectories satisfying the boundary
conditions (\ref{bound}) should be included in the semiclassical
propagator.  The trajectories for which the real part of the exponent
$\phi$ in (\ref{smivrag}) is positive must be discarded as they give
rise to divergent contributions in the semiclassical limit. These
trajectories are probably associated with forbidden deformations of the
integration contours that are necessary to derive the semiclassical
approximation (\ref{csp}).

For the harmonic oscillator it can be checked explicitly that not only
equations (\ref{smivrg}) and (\ref{smivrag}) give exact results but
also that $Re(\phi) \leq 0$ for all complex trajectories. In our
calculations trajectories satisfying
\begin{equation}
Re(\phi) > c \hbar,
\label{cutoff}
\end{equation}
where $c$ is a constant, are neglected. We discuss the importance of
the cutoff value of $c$ in the next section.

%%%%%%%%%%%%%%%%%%%%%%%%%%%%%%%%%%%%%%%%%%%%%%%%%%%%%%%%%%%%%%%%%%%%%
%%%%%%%%%%%%%%%%%%%%%%%%%%%%%%%%%%%%%%%%%%%%%%%%%%%%%%%%%%%%%%%%%%%%%
\section{Example}
\label{sec_Ex}

As a simple application of the smooth CIVR we consider the system
\begin{equation}
\hat{H} = \frac{1}{2}\hat{p}^2 + \frac{\Omega^2}{2} \hat{q}^2 +
\frac{\lambda}{4} \hat{q}^4.
\end{equation}
It has been studied also in \cite{Agu05} by directly computing the
relevant complex trajectories. The parameters are set to $\Omega=1$,
$\lambda=0.4$ and $\hbar=1\;.$ For these values the ground state energy
is $E_0\approx 0.559$ and the first two excited states have $E_1\approx
1.770$ and $E_2\approx 3.319$. For the initial wavepacket we choose
$q_0=0$, $p_0=-2.0$, and $b=1.0$. This gives $E=H(q,p)=2.0$ for the
energy of the central trajectory, $\tau \approx 4.7$ for its period,
and $X_{\rm{turn}} \approx \pm 1.6$ for its turning points. Figure 1(c)
shows a plot of the potential function and indicates also
the central trajectory energy.

We momentarily restore the original un-scaled variables to illustrate both
the computation of the classical Hamiltonian and the scaling process.
The classical Hamiltonian function is
\begin{equation}
H = \frac{1}{2}p^2 + \frac{1}{2} \left(\Omega^2+\frac{3\lambda b^2}{4}\right) q^2 +
\frac{\lambda}{4} q^4 + \left(\frac{\hbar^2}{4b^2} +
\frac{\Omega^2 b^2}{4}+\frac{3\lambda b^4}{16}\right),
\end{equation}
where $b$ is the width of the wavepacket. In terms of scaled variables
(see equation (\ref{barvar})) the Hamiltonian becomes
\begin{equation}
\bar{H} = \omega \left[\frac{1}{2}\bar{p}^2 + \frac{1}{2} \bar{\nu}^2 q^2 +
\frac{\bar{\lambda}}{4} \bar{q}^4 + \frac{1}{4}\left(1 + \nu^2+
\frac{3\bar{\lambda}}{16}\right) \right],
\end{equation}
where $\omega= \hbar/b^2$, $\nu=\Omega/\omega$, $\bar{\lambda} =
\lambda \hbar/\omega^3$ and $\bar{\nu}^2= \nu^2 + 3\bar{\lambda}/2$.
For the present values we have
$\omega=\nu=1$, $\bar{\lambda}=0.4$ and
$\bar{\nu}^2= 1.6$.

Figure 1 shows five snapshots of the wavepacket (left column) and the
corresponding regions of the $q_1,p_1$ plane where trajectories
contribute significantly to the propagation. In these figures we have
fixed the constant $c=1.0$ (see equation (\ref{cutoff})), except for
figure 1(a), where $c=2.5$. The width $a$ of the smoothing Gaussian was
adjusted to get the best results for each propagation time, starting at
$a=1.5$ for $T=1.0$ and decreasing to $a=0.4$ for $T=8.5$ (see caption
for all values). The integration over $q_1$ and $p_1$ was performed
using a regular grid with 30 points in $q_1$, varying from $-3$ to $3$,
and 40 points in $p_1$ varying from $-4$ to $4$. The computational time
for the present calculation is as fast as the split-operator method,
well known for being efficient and accurate for one dimensional
problems. For $T=8.5$ the calculations take about 3 seconds in a Core 2
Quad PC with 2.4GHz.

The wavefunctions in figure 1 were calculated using the simple
discretization
\begin{equation}
\psi(x,T) = \sum_{n,m} \langle x|z_{nm}\rangle K(z_{nm}^*,z_0,T)
\frac{\Delta q \Delta p}{2\pi}
\end{equation}
where $n$ and $m$ represent the grid in phase-space centered on the
origin. We used a total of 40 and 60 points in the $q$ and $p$
directions respectively, with $-4 < q_n < +4$ and $-6 < p_m < +6$.

In spite of the accuracy of our results, specially as compared to
previous calculations using root search procedures \cite{Agu05},
several details remain to be understood and improved. The main problem
is the sensitivity of the method to the choice of the width $a$ and the
lack of a theory on how to choose it properly and automatically. A
possible way out of this difficulty might to be the procedure devised
in \cite{Mak87}, where the width is chosen to minimize the oscillations
of the integrand. Another problem is that the propagated wavepackets
turn out not to be properly normalized, and the amount by which
normalization is lost also depends on the width $a$. In figure 1 the
wave-functions have been re-normalized by hand after the propagation.

Despite these problems the method improves the results obtained by
direct computation of the contributing trajectories and is much faster
and simple to program. The next step is an application of the method to
multidimensional systems, where the integrations over the initial
conditions may be performed by Monte Carlo techniques. The difficulties
just mentioned are currently under investigation.

\appendix
%%%%%%%%%%%%%%%%%%%%%%%%%%%%%%%%%%%%%%%%%%%%%%%%%%%%%%%%%%%%%%%%%%%%%
%%%%%%%%%%%%%%%%%%%%%%%%%%%%%%%%%%%%%%%%%%%%%%%%%%%%%%%%%%%%%%%%%%%%%
\section{Tangent matrices}
\label{appA}

In this appendix we use the scaled units where $\hbar=b=1$. In the $u$
and $v$ variables the tangent matrix is defined by
\begin{equation}
\left( \begin{array}{l}
    \delta u(T) \\
    \delta v(T) \end{array} \right) =
\left( \begin{array}{cc}
    M_{uu} \quad &  M_{uv}  \\
    M_{vu} \quad &  M_{vv}
    \end{array} \right)
\left( \begin{array}{l}
    \delta u(0) \\
    \delta v(0) \end{array} \right)
\label{a1}
\end{equation}
where $\delta u(0)$ and $\delta v(0)$ are small displacements at the
initial point of the trajectory and $\delta u(T)$ and $\delta v(T)$ are
the corresponding final deviations. The action $S(v'',u',T)$ for the
trajectory with $u(0)=u'$ and $v(T)=v''$ satisfies \cite{Bar01}
\begin{equation}
\label{a2}
u(T) \equiv u'' = i \frac{\partial S}{\partial v''},\qquad
v(0) \equiv v'= i \frac{\partial S}{\partial u'}.
\end{equation}
From the differentiation of  (\ref{a2}) keeping the variable $T$
constant, we can obtain the connection between initial and final
displacements. In matrix form it is
\begin{equation}
\left( \begin{array}{l}
    \delta u(T) \\
    \delta v(0) \end{array} \right) =
i \left( \begin{array}{cc}
    S_{uu} \quad &  S_{uv}  \\
    S_{vu} \quad &  S_{vv}
    \end{array} \right)
\left( \begin{array}{l}
    \delta u(0) \\
    \delta v(T) \end{array} \right)
\label{a3}
\end{equation}
where $S_{uv} = \partial^2 S/\partial u' \partial v''$, etc. Comparing
with eq.(\ref{a1}) we find
\begin{equation}
S_{uv} = -i M_{vv}^{-1} \qquad \qquad
S_{vv} = -i \frac{M_{uv}}{M_{vv}}.
\label{a4}
\end{equation}

Using the definition of $u$ and $v$ in terms of $q$ and $p$ (notice
that all these variables are complex) it is easy to show that
\cite{Bar01}
\begin{equation}
\begin{array}{ll}
M_{uu} &= \frac{1}{2}(m_{qq}+m_{pp}+i m_{pq}-i m_{qp}) \\
M_{uv} &= \frac{1}{2}(m_{qq}-m_{pp}+i m_{pq}+i m_{qp}) \\
M_{vu} &= \frac{1}{2}(m_{qq}-m_{pp}-i m_{pq}-i m_{qp}) \\
M_{vv} &= \frac{1}{2}(m_{qq}+m_{pp}-i m_{pq}+i m_{qp}) ,
\end{array}
\end{equation}
where $m$ is the tangent matrix in the $q$, $p$ system. Finally, using
the definition of the real variables $Q_1$, $Q_2$, $P_1$, $P_2$ and
defining its corresponding $4 \times 4$ tangent matrix $n$ we can show
that
\begin{equation}
\begin{array}{ll}
m_{qq} &= n_{11}-in_{14} \\
m_{qp} &= n_{13}-in_{12} \\
m_{pq} &= n_{24}+in_{21} \\
m_{pp} &= n_{22}+in_{23} .
\end{array}
\end{equation}
Therefore, by working directly with the real trajectories in the double
phase space we can compute $n$ and reconstruct the matrices $m$  and
$M$ using simple linear transformations.

%%%%%%%%%%%%%%%%%%%%%%%%%%%%%%%%%%%%%%%%%%%%%%%%%%%%%%%%%%%%%%%%%%%%%
%%%%%%%%%%%%%%%%%%%%%%%%%%%%%%%%%%%%%%%%%%%%%%%%%%%%%%%%%%%%%%%%%%%%%
\section{Calculation of $\det{\Lambda}$}
\label{appB}

If $v(T)$ in equation (\ref{eqvt}) is an analytic function of the
initial condition $v_1$, then, by the Cauchy-Riemann conditions we have
\begin{equation}
\dps{\frac{\partial q_1(T)}{\partial q_1} = \frac{\partial p_1(T)}{\partial p_1},
\qquad \qquad
\frac{\partial q_1(T)}{\partial p_1} = -\frac{\partial p_1(T)}{\partial q_1}}.
\label{b1}
\end{equation}
By the definition of $\Lambda$, equation (\ref{civr6}),
\begin{equation}
\det{\Lambda} = \left(\frac{\partial q_1(T)}{\partial q_1}\right)^2 +
\left(\frac{\partial q_1(T)}{\partial p_1}\right)^2.
\label{b2}
\end{equation}
On the other hand we also have,
\begin{equation}
\begin{array}{ll}
\dps{\frac{\partial v(T)}{\partial v_1}}   & =
\dps{ \frac{1}{\sqrt{2}}
\left( \frac{\partial }{\partial q_1}+ i \frac{\partial }{\partial p_1}\right)
\frac{1}{\sqrt{2}}
\left( q_1(T) - i p_1(T)\right)} \\
& = \dps{\frac{1}{2}\left( \frac{\partial q_1(T)}{\partial q_1} +
\frac{\partial p_1(T)}{\partial p_1}\right) +
\frac{i}{2}\left( \frac{\partial q_1(T)}{\partial p_1} -
\frac{\partial p_1(T)}{\partial q_1}\right)}\\
& = \dps{\frac{\partial q_1(T)}{\partial q_1} + i \frac{\partial q_1(T)}{\partial p_1}}
\end{array}
\label{b3}
\end{equation}
and, therefore,
\begin{equation}
\det{\Lambda} = \dps{\left|\frac{\partial v(T)}{\partial v_1}\right|^2}.
\label{b4}
\end{equation}
Finally, using the second of equations (\ref{a2}) with $v'=v_1$, and
differentiating with respect to $v(T)$,
\begin{equation}
\frac{\partial v_1}{\partial v(T)}= i \frac{\partial^2 S}{\partial u' \partial v(T)}
\label{b5}
\end{equation}
which implies
\begin{equation}
\det{\Lambda} = \dps{\left|i \frac{\partial^2 S}{\partial u' \partial v(T)} \right|^{-2}} =
|M_{vv}|^2
\label{b6}
\end{equation}
by equation (\ref{a4}).

%%%%%%%%%%%%%%%%%%%%%%%%%%%%%%%%%%%%%%%%%%%%%%%%%%%%%%%%%%%%%%%%%%%%%
%%%%%%%%%%%%%%%%%%%%%%%%%%%%%%%%%%%%%%%%%%%%%%%%%%%%%%%%%%%%%%%%%%%%%
\section{Acknowledgements}

Support from FAPESP and CNPq (Brazil) is acknowledged.
Facilities of the CENAPAD high-performance computing center
at Universidade Estadual de Campinas where used in this work.

\newpage

\begin{figure}[h!] \centering
\includegraphics[width=.35\textwidth,angle=-90]{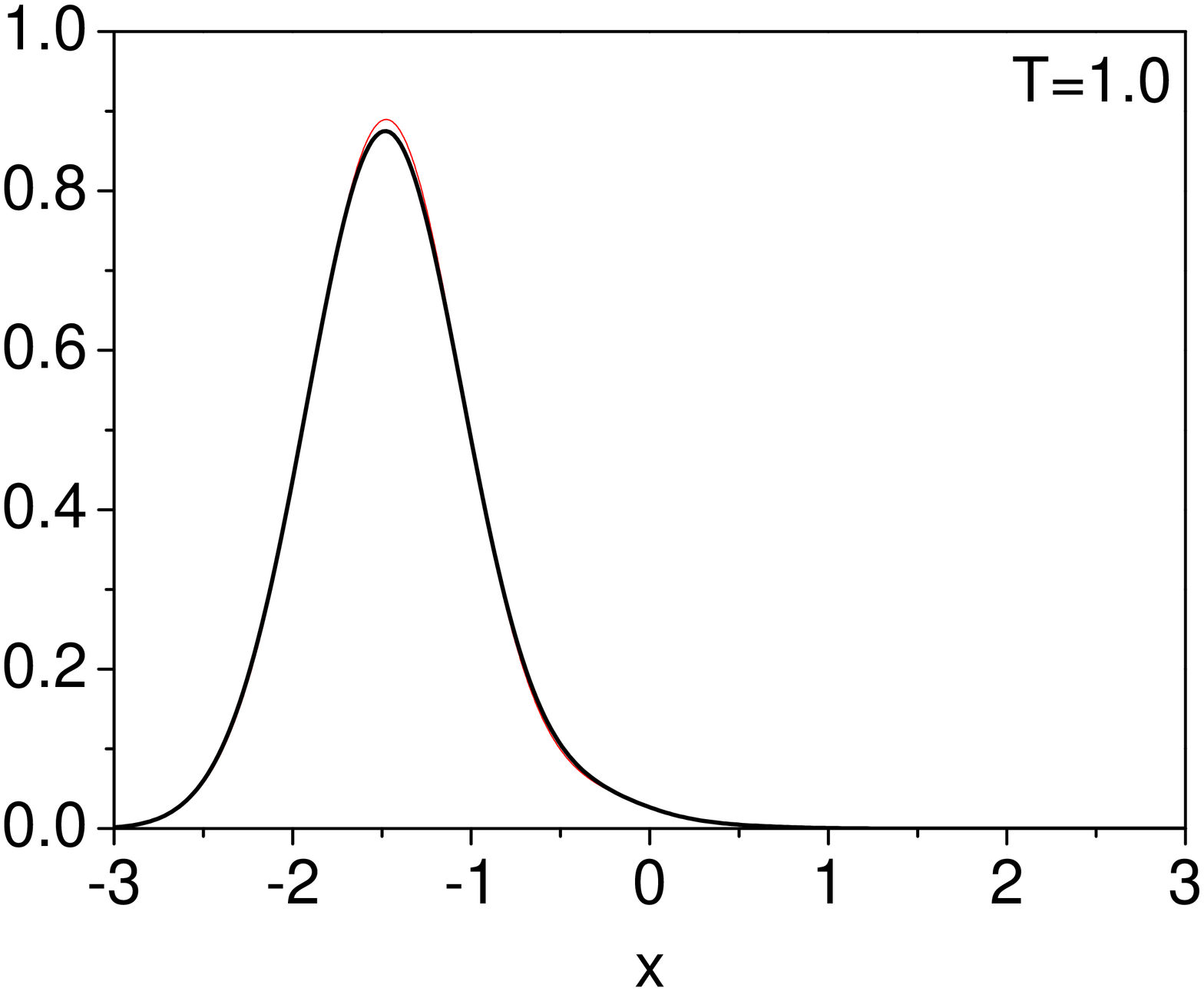}
\includegraphics[width=.35\textwidth,angle=-90]{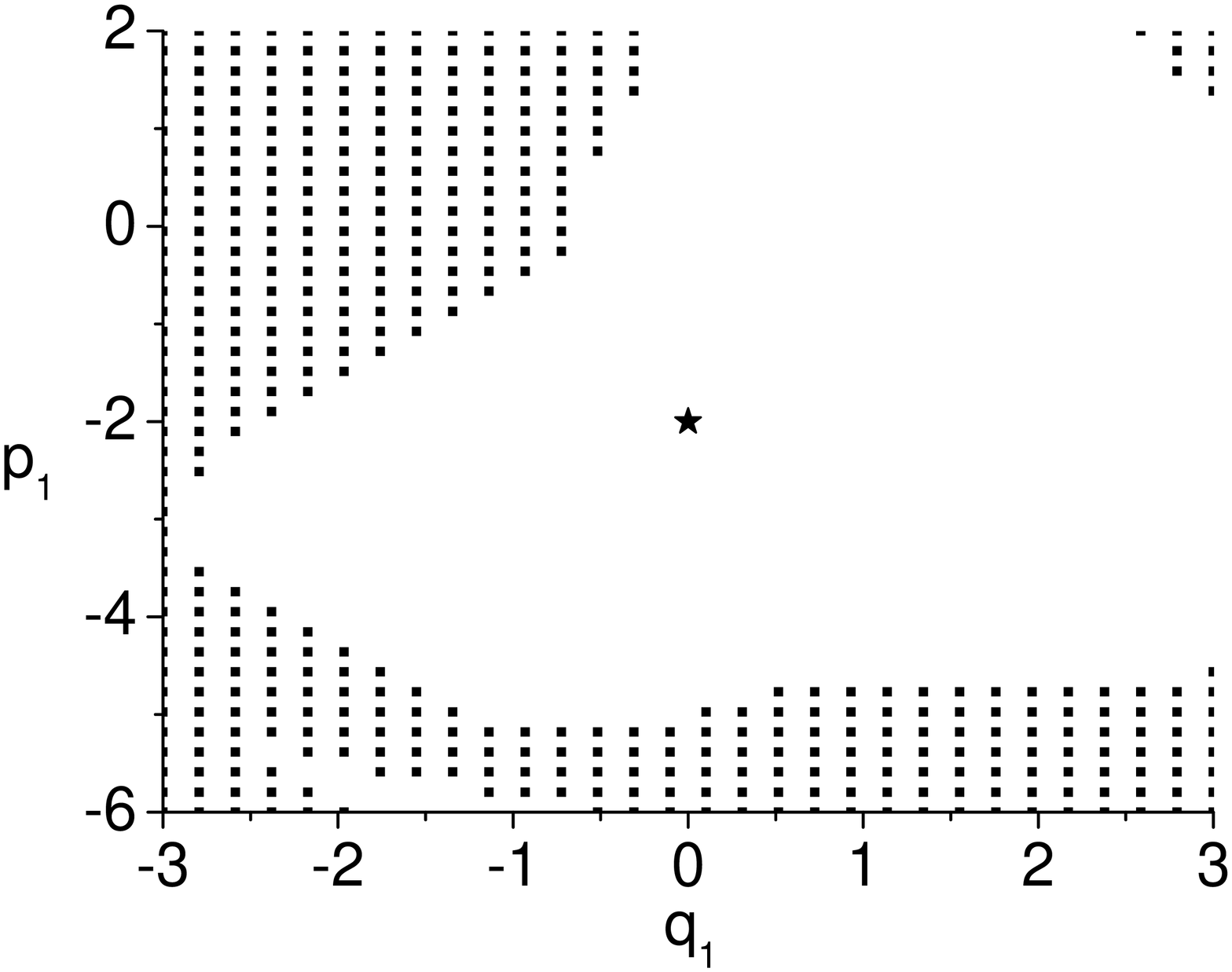}
\includegraphics[width=.35\textwidth,angle=-90]{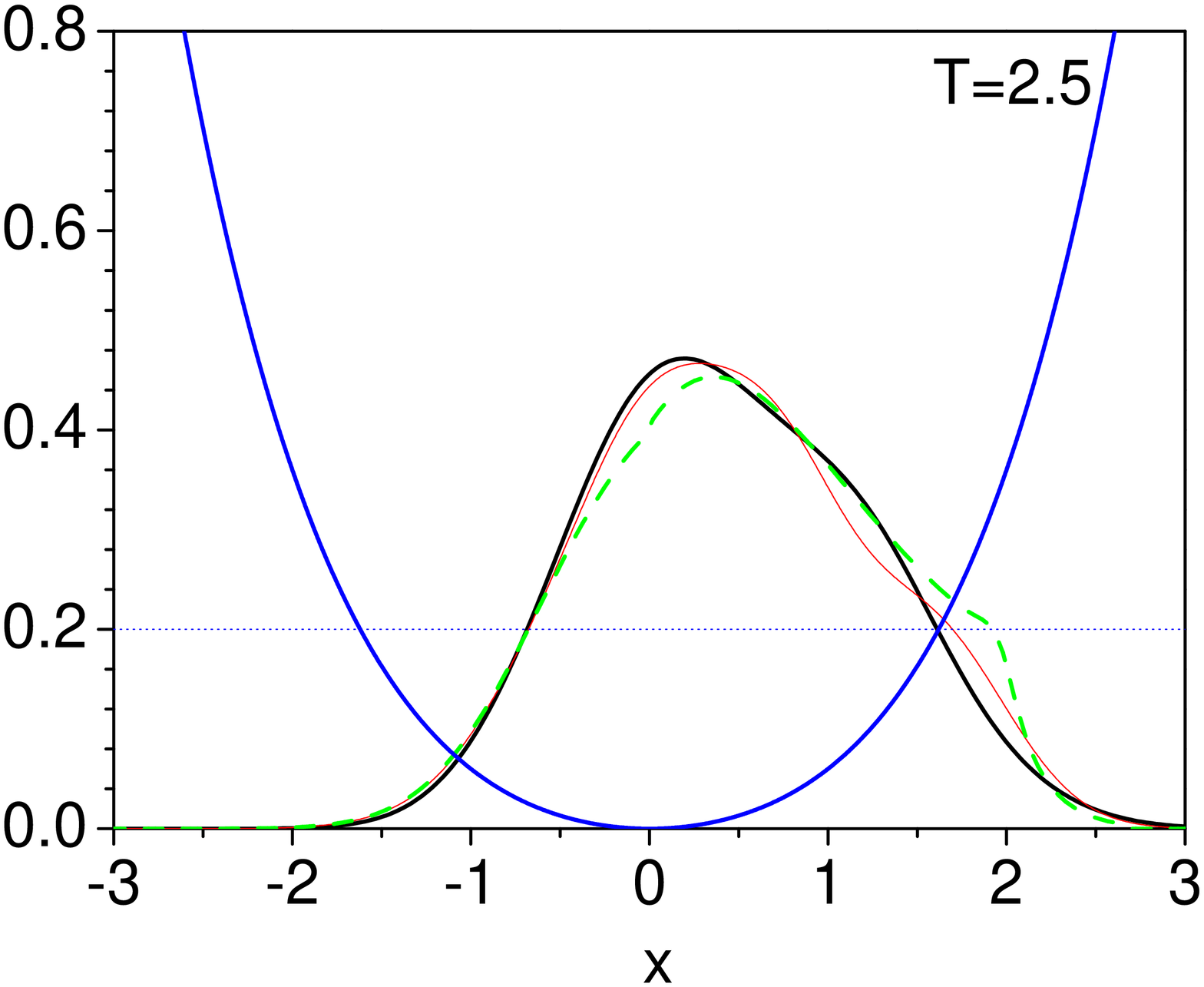}
\includegraphics[width=.35\textwidth,angle=-90]{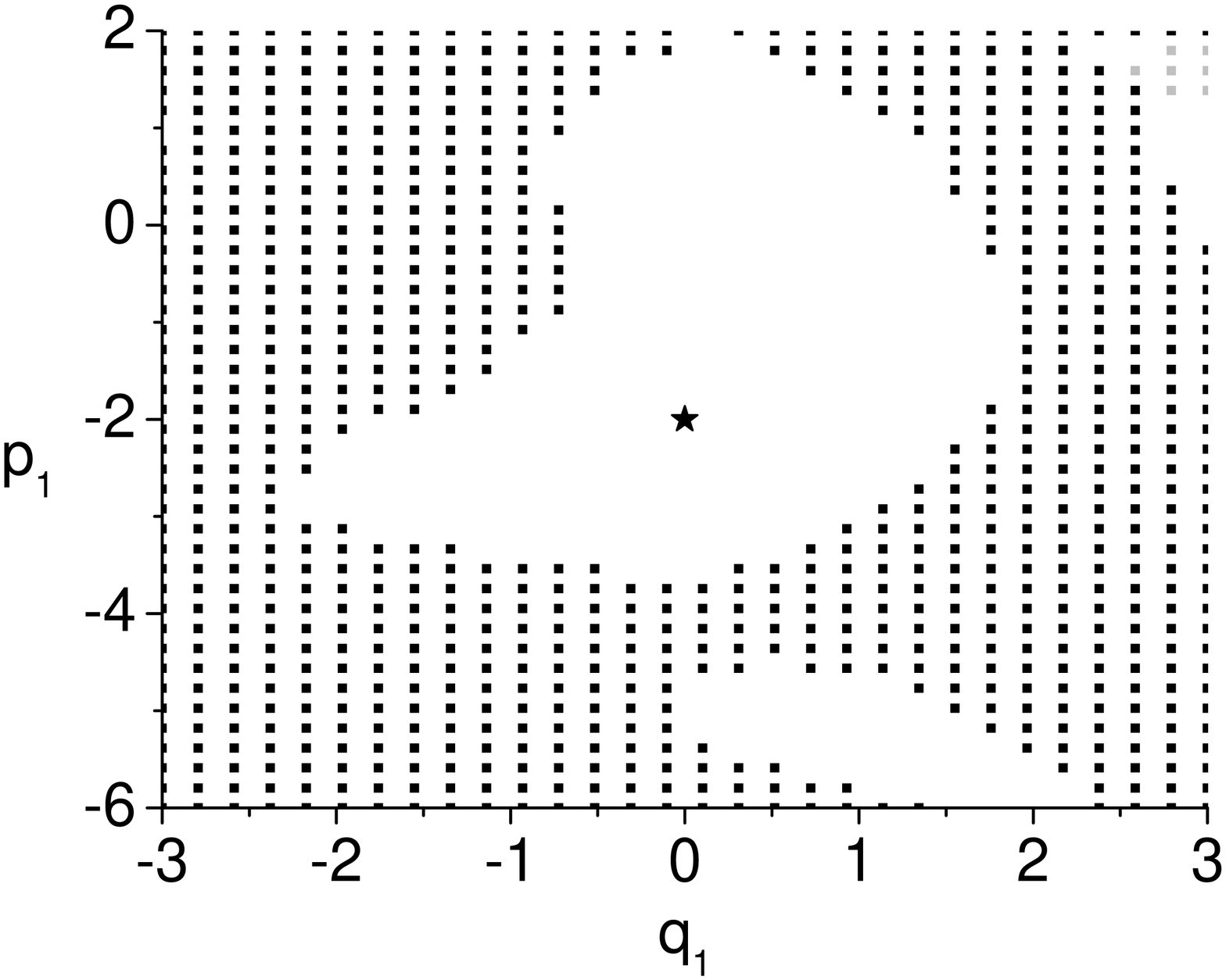}
\includegraphics[width=.35\textwidth,angle=-90]{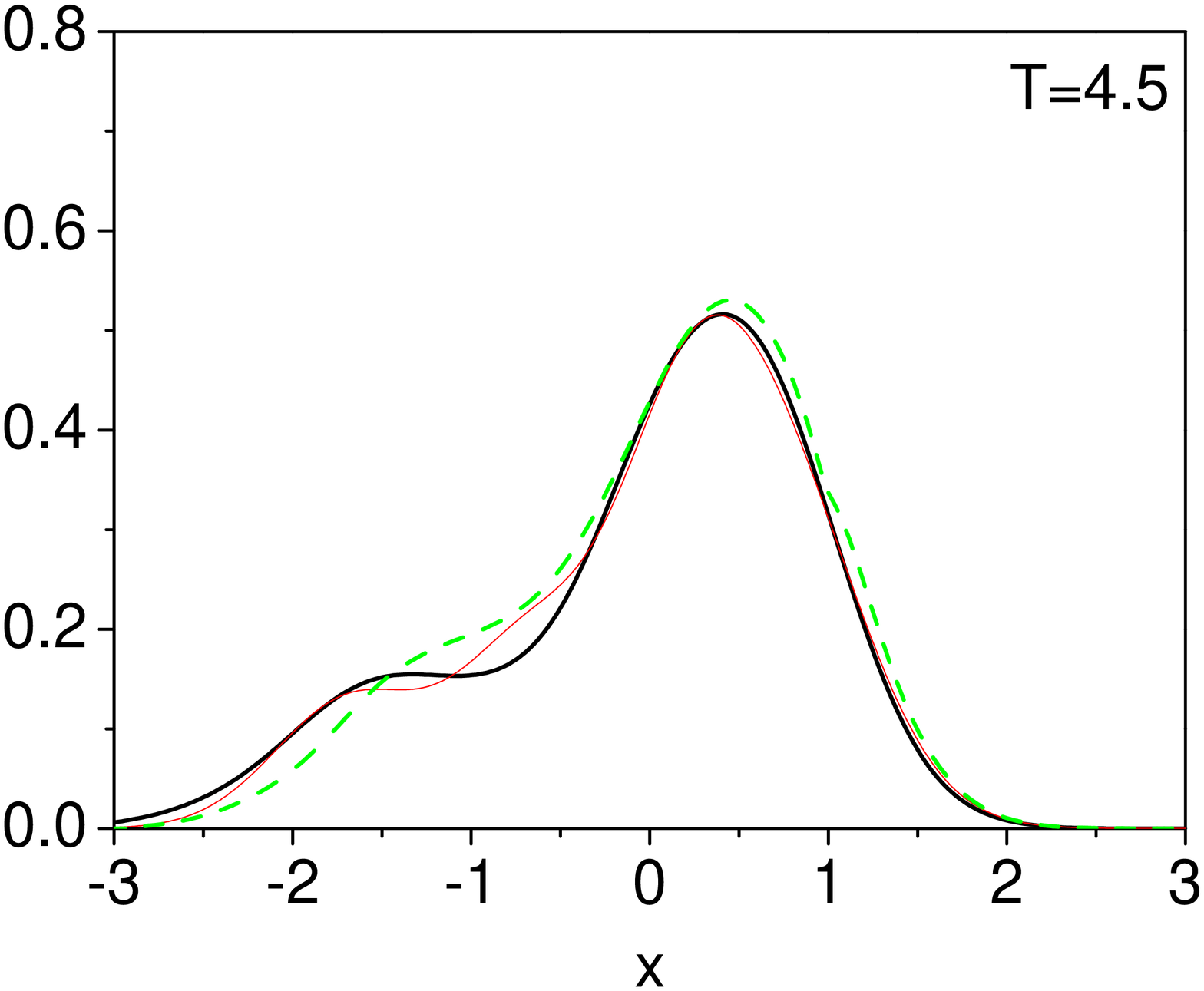}
\includegraphics[width=.35\textwidth,angle=-90]{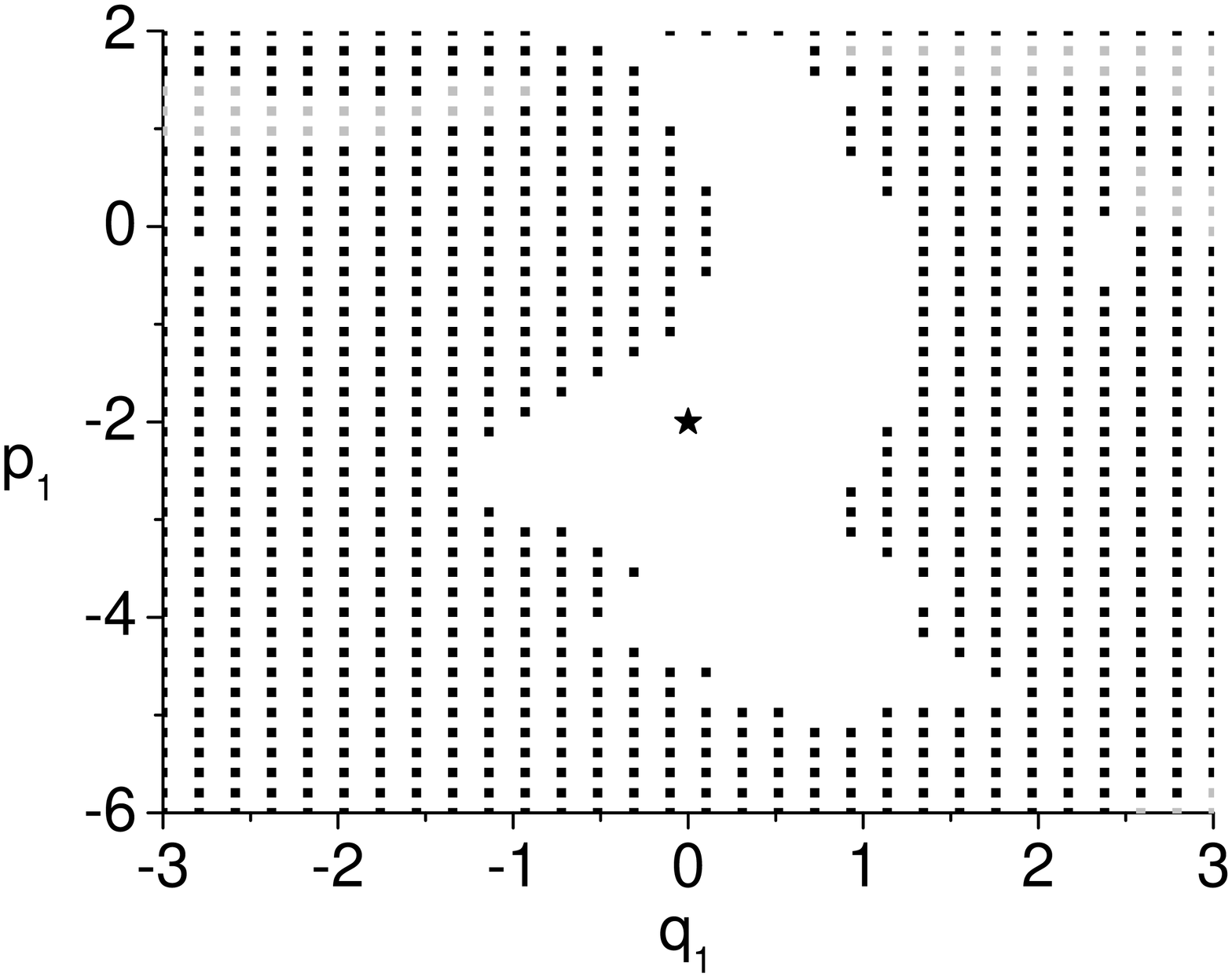}
\includegraphics[width=.35\textwidth,angle=-90]{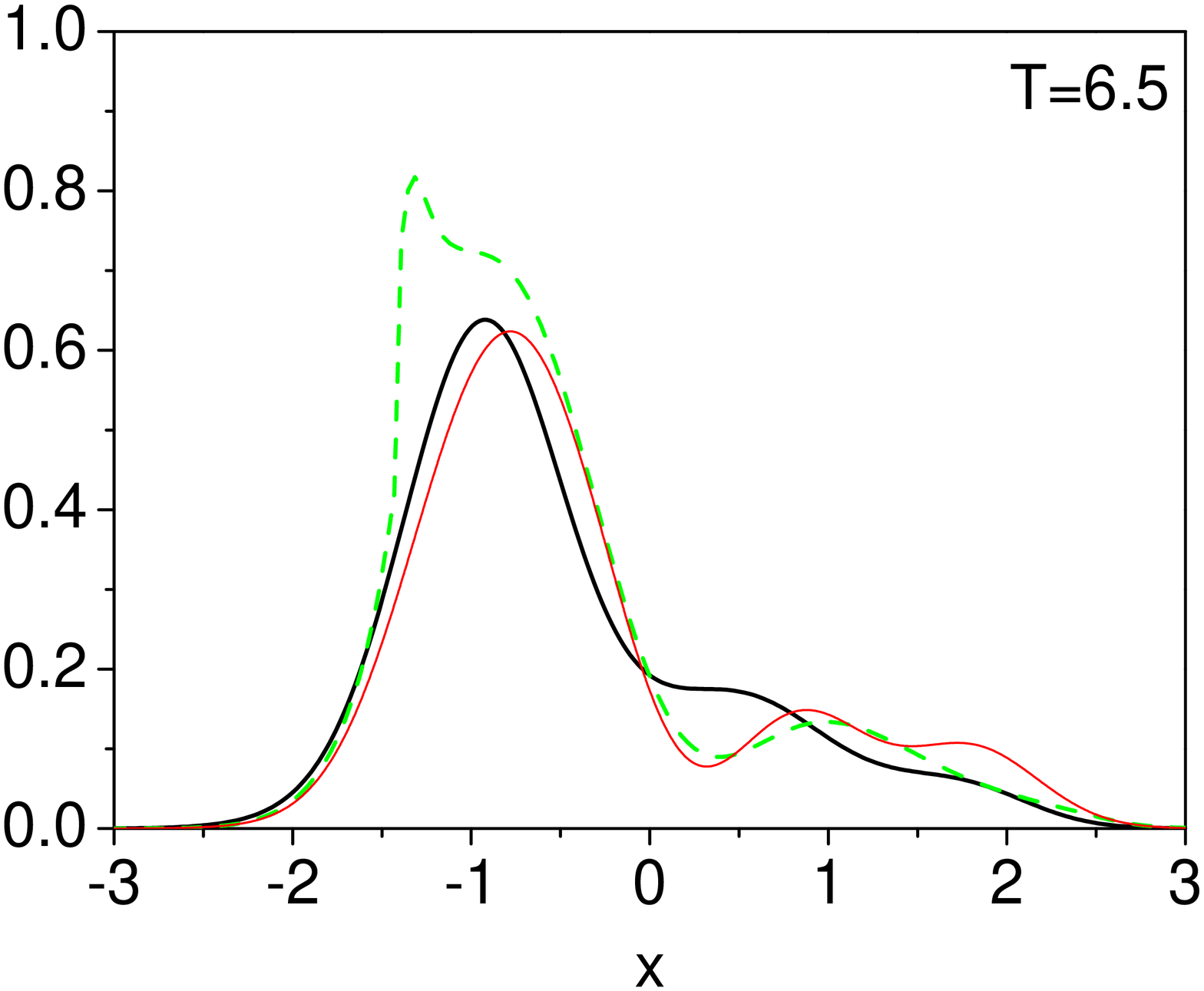}
\includegraphics[width=.35\textwidth,angle=-90]{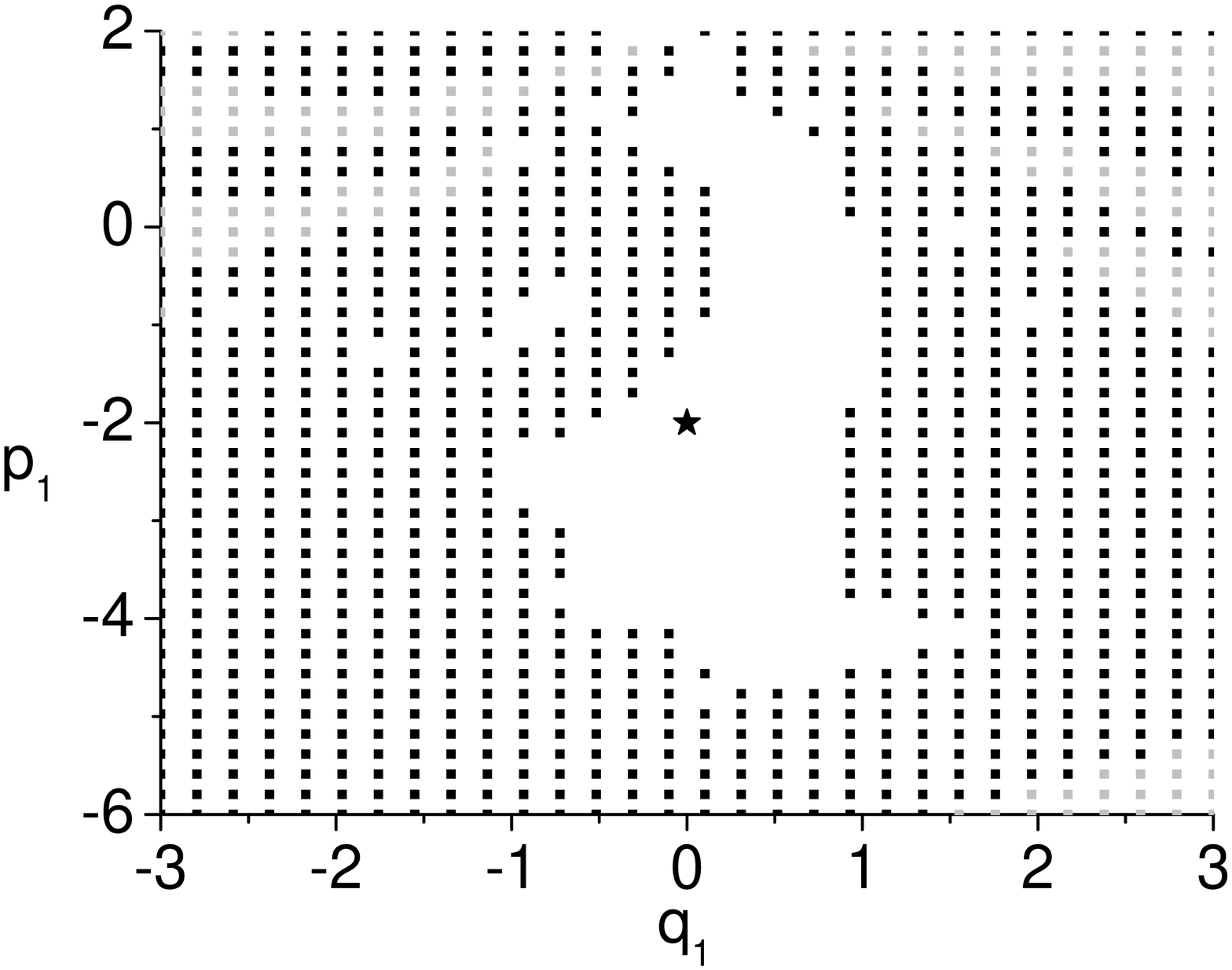}
\includegraphics[width=.35\textwidth,angle=-90]{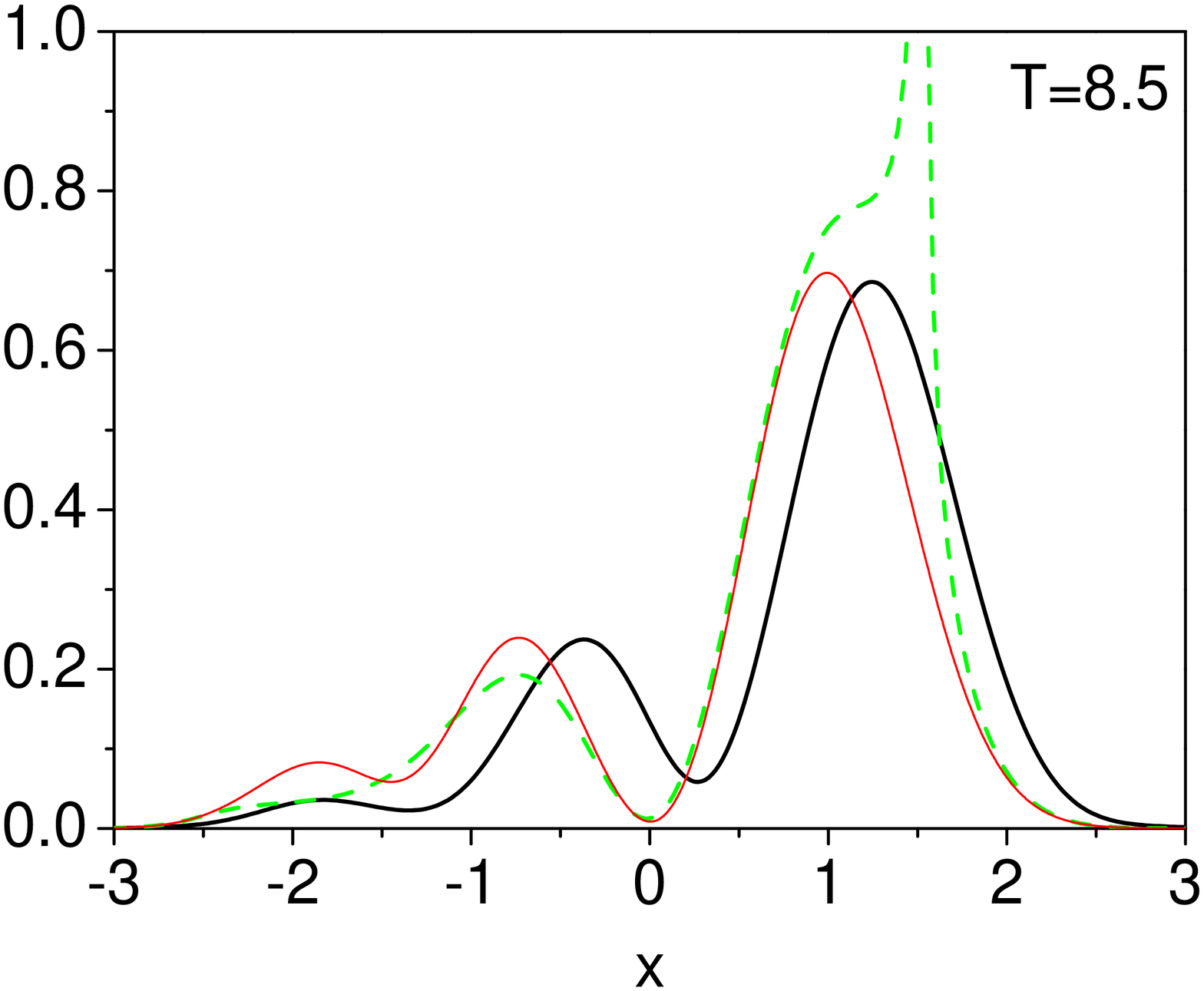}
\includegraphics[width=.35\textwidth,angle=-90]{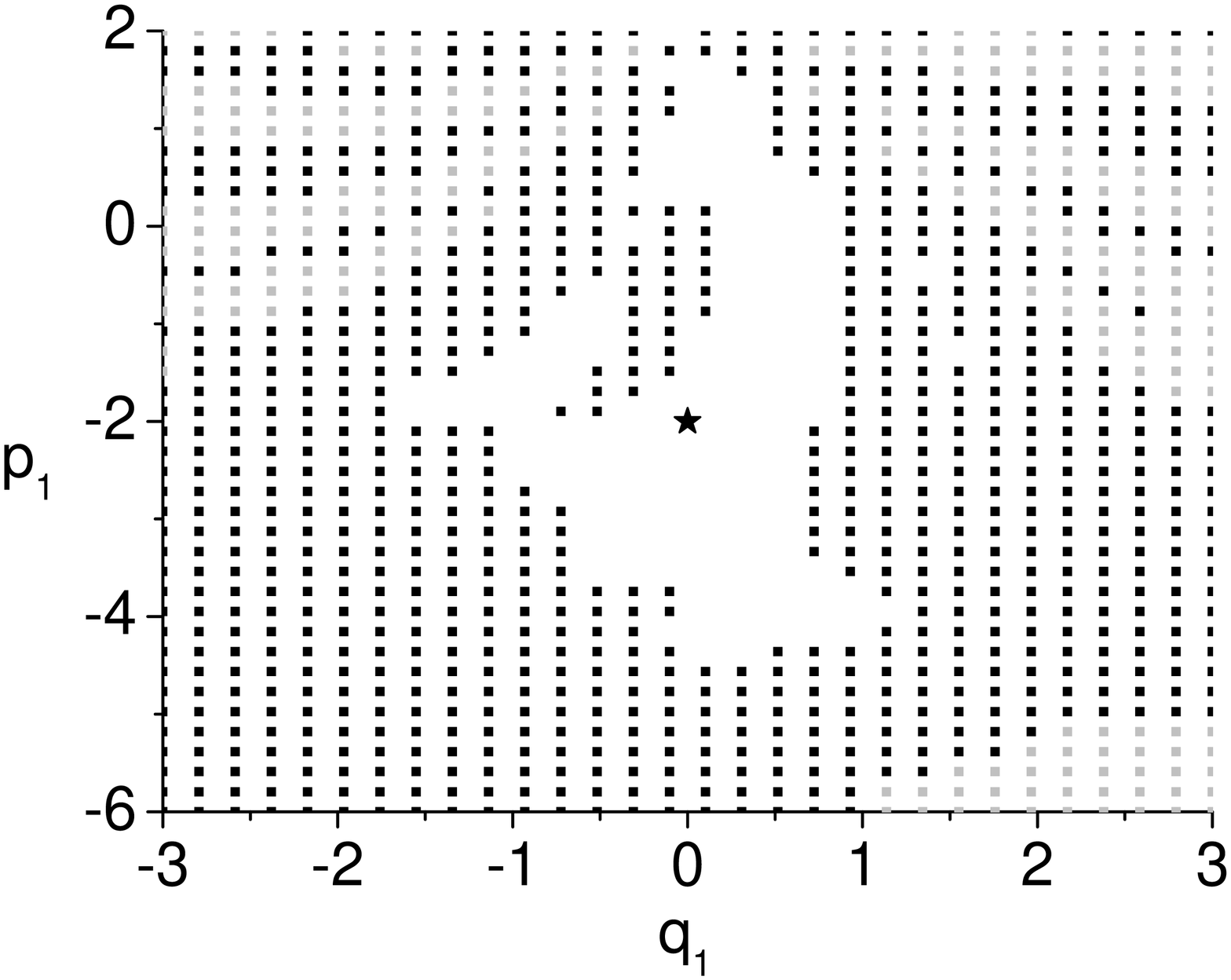}
\label{fig1}
\end{figure}

\newpage

Figure Caption:\\

Figure 1. (color online) The right panels show the exact and
semiclassical wavefunctions for several values of $T$. The thin
continuous line (red) displays the exact result obtained via split
operator method; the thick solid line is the CIVR approximation and the
dashed line (green) shows the result obtained in ref.\cite{Agu05} by
direct computation of contributing trajectories. For $T=2.5$ we also
show the potential ($V(x)/10$) and the energy $E=2.0$ of the central
trajectory (shown as $E/10$). The left panel shows the contributing and
non-contributing initial trajectories as white and dark areas
respectively. The star indicates the positions $q_0$ and $p_0$ of the
initial wavepacket.

\end{document}